\documentclass[
 reprint,
 groupedaddress,
 amsmath,amssymb,
 aip,
 apl,
]{revtex4-1}

\usepackage{float}
\usepackage{graphicx}
\usepackage{dcolumn}
\usepackage{bm}

\usepackage{xcolor}
\usepackage[normalem]{ulem}
\usepackage[colorlinks=true,allcolors=blue]{hyperref}

\definecolor{commentColor1}{rgb}{0.8,0.0,0.2}
\definecolor{commentColor2}{rgb}{0.0,0.2,0.6}
\definecolor{commentColor3}{rgb}{0.6,0.2,0.0}

\newcommand{\omegap}{\omega_\text{p}}
\newcommand{\omegapt}{\tilde{\omega}_\text{p}}
\newcommand{\omegao}{\omega_0}
\newcommand{\omegaot}{\tilde{\omega}_0}

\newcommand{\omegar}{\omega_{\text{r}}}

\newcommand{\kappae}{\kappa_\text{e}}
\newcommand{\phizpf}{\phi_\text{zpf}}

\newcommand{\Cone}{C_1}
\newcommand{\Cin}{C_\text{in}}
\newcommand{\Lone}{L_1}

\newcommand{\Czero}{C_0}
\newcommand{\Cm}{C_\text{m}}
\newcommand{\Lm}{L_\text{m}}
\newcommand{\Rm}{R_\text{m}}

\newcommand{\Ksq}{K^2}

\newcommand{\Cg}{C_\text{g}}

\newcommand{\Z}{Z_0}

\newcommand{\Qi}{Q_{i}}
\newcommand{\Qires}{Q_{i,\text{res}}}
\newcommand{\Qirel}{Q_{i,\text{rel}}}
\newcommand{\Qe}{Q_{e}}
\newcommand{\deltaTLS}{\delta_{\text{TLS}}^0}
\newcommand{\massflux}{\Phi_{\text{m}}}
\newcommand{\meff}{m_\text{eff}}

\begin{document}

\title{Loss channels affecting lithium niobate phononic crystal resonators at cryogenic temperature}

\author{E. Alex Wollack}
\thanks{Authors to whom correspondence should be addressed: \href{mailto:ewollack@stanford.edu}{ewollack@stanford.edu}, \href{mailto:safavi@stanford.edu}{safavi@stanford.edu}}
\author{Agnetta Y. Cleland}
\author{Patricio Arrangoiz-Arriola}
\thanks{Present address: {\it AWS Center for Quantum Computing, Pasadena, California, USA}}
\author{Timothy P. McKenna}
\author{Rachel G. Gruenke}
\author{Rishi N. Patel}
\author{Wentao Jiang}
\author{Christopher J. Sarabalis}
\author{Amir H. Safavi-Naeini}
\thanks{Authors to whom correspondence should be addressed: \href{mailto:ewollack@stanford.edu}{ewollack@stanford.edu}, \href{mailto:safavi@stanford.edu}{safavi@stanford.edu}}
\affiliation{Department of Applied Physics and Ginzton Laboratory, Stanford University\\348 Via Pueblo Mall, Stanford, California 94305, USA}

\date{\today}

\begin{abstract}
We investigate the performance of microwave-frequency phononic crystal resonators fabricated on thin-film lithium niobate for integration with superconducting quantum circuits. For different design geometries at millikelvin temperatures, we achieve mechanical internal quality factors $Q_i$ above $10^5 - 10^6$ at high microwave drive power, corresponding to $5\times10^6$ phonons inside the resonator. By sweeping the defect size of resonators with identical mirror cell designs, we are able to indirectly observe signatures of the complete phononic bandgap via the resonators' internal quality factors. Examination of quality factors' temperature dependence shows how superconducting and two-level system (TLS) loss channels impact device performance. Finally, we observe an anomalous low-temperature frequency shift consistent with resonant TLS decay and find that material choice can help to mitigate these losses.
\end{abstract}

\maketitle

The field of circuit quantum acousto-dynamics (cQAD) has recently gained traction as a viable way to achieve quantum control of mechanical resonators via piezoelectric coupling to a superconducting qubit ~\cite{OConnell2010,Arrangoiz-Arriola2016,Chu2017,Chu2018,Satzinger2018,Arrangoiz-Arriola2019,Sletten2019,Chu2020}. Manipulation and measurement of the quantum states of bulk acoustic wave (BAW)~\cite{Chu2017,Chu2018}, surface acoustic wave (SAW)~\cite{Satzinger2018,Sletten2019}, and phononic crystal~\cite{Arrangoiz-Arriola2019} resonators have demonstrated the capability of these hybrid systems. One of the primary motivations for integrating acoustic resonators with superconducting qubits is the advantages that mechanical systems can offer for scaling qubit architectures~\cite{Ofek2016,Lescanne2020} that utilize bosonic systems to store quantum information. Not only have mechanical resonators been shown to have lifetimes comparable to~\cite{Renninger2018} and exceeding~\cite{MacCabe2019} conventional superconducting resonators~\cite{Reagor2016,Megrant2012}, but they can also be extremely compact due to the slow speed of sound in most materials. Combined with proposals utilizing mechanical resonators as quantum memories~\cite{Pechal2018,Hann2019}, these systems offer a path towards increasing computational complexity of existing superconducting qubit processors.

Towards achieving these goals, phononic crystal resonators offer some unique advantages over other cQAD platforms. The complete phononic bandgap induced by the crystal's periodic patterning not only tightly localizes mechanical motion, but also offers protection against spurious acoustic radiation of the superconducting circuit. In contrast to approaches based on bulk acoustic or surface acoustic waves~\cite{Scigliuzzo2020}, the phononic bandgap suppresses scattering loss due to inevitable fabrication disorder, such as surface or electrode roughness or isotopic impurities~\cite{Kostylev2017}. It is also straightforward to precisely engineer the frequency and coupling strength of these resonators, a key requirement to be able to perform quantum operations without crosstalk in certain proposals~\cite{Hann2019}. Despite these advantages, recent demonstrations of coupling phononic crystal resonators to qubits have been limited by the coherence times of the mechanical system~\cite{Arrangoiz-Arriola2019}, motivating the need to better understand these devices' performance and limitations.

\begin{figure*}[t]
    \centering
    \includegraphics[width=170mm]{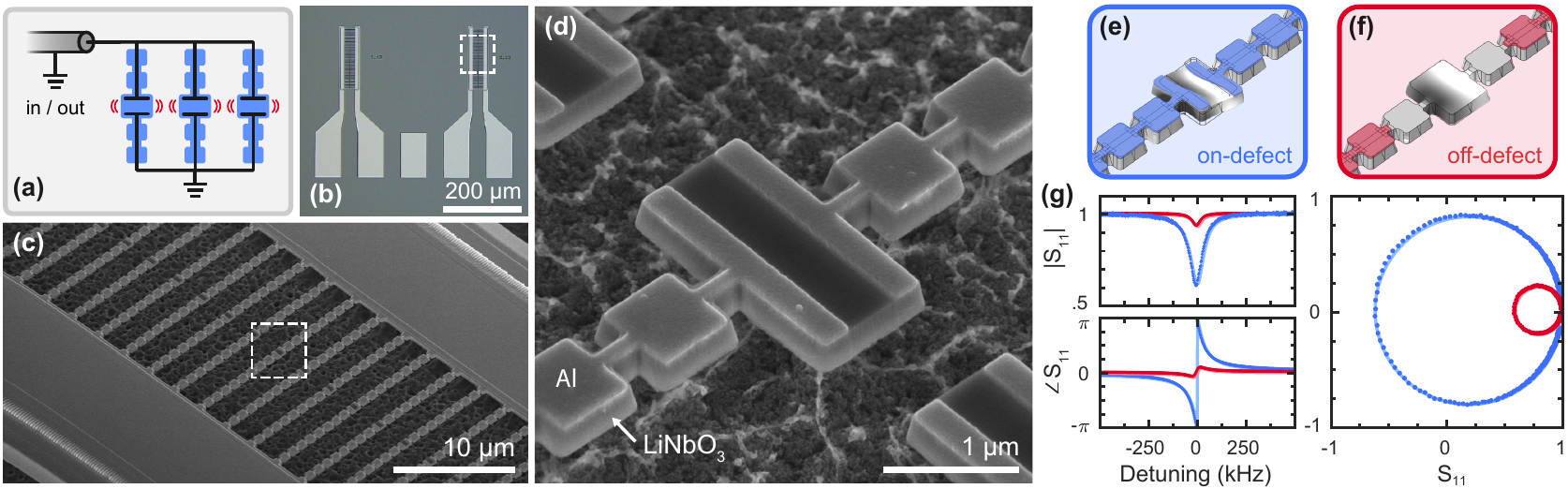}
    \caption{{\bf Device overview.}  (a)~Schematic of the device, with phononic crystal resonators inserted at the end of a microwave transmission line and measured on reflection. (b)~Optical micrograph of two neighboring devices, each with 40 mechanical resonators placed between aluminum contact pads. (c)~Scanning-electron micrograph (SEM) of the suspended mechanical resonators. (d)~SEM image of the phononic crystal defect site showing aluminum electrodes on top of lithium niobate.  (e,f)~Finite-element simulations of the defect site's strongly-coupled mode, showing the structure's localized mechanical deformation and associated electrostatic potential (greyscale). The aluminum electrodes (shaded blue/red) extend onto the defect site for the on-defect designs (e) in contrast to the off-defect designs (f), where the electrodes are one unit cell removed from the mechanically active defect region. (g) Normalized reflection $S_{11}$ measurements for on-defect (blue) and off-defect (red) resonators, with fits to Eq.~\ref{eq_S11} (solid lines). The magnitude $|S_{11}|$ and phase $\angle S_{11}$ are plotted versus the detuning relative to the resonant frequency $\omega_r/2\pi \simeq 2.0\,\text{GHz}$. $S_{11}$ is also plotted in the complex plane (horizontal axis: real part, vertical axis: imaginary part). The response of the off-defect device is enlarged by a factor of $\sim10$ for visual clarity.}
    \label{fig_device}
\end{figure*}

The nanomechanical resonators explored in this study are one-dimensional phononic crystal resonators made from thin-film lithium niobate (LN), each consisting of a defect site embedded in an acoustic shield (Fig.~\ref{fig_device}(a-d))~\cite{Arrangoiz-Arriola2016,ArrangoizArriola2018,Arrangoiz-Arriola2019}. The periodic patterning of the shield opens a complete phononic bandgap for all acoustic polarizations in the $1.65-2.1\,\text{GHz}$ range, confining mechanical motion to the defect and thereby minimizing clamping losses. Due to lithium niobate's strong piezoelectricity, aluminum electrodes placed on or near the defect are able to achieve large electromechanical coupling to the fundamental shear mode of the defect site. By sweeping the size of the defect width from $0.5$ to $2.5\,\mu\text{m}$, a large array of devices is densely frequency multiplexed from $1.5-3.0\,\text{GHz}$ to cover frequencies both inside and outside of the phononic bandgap. In studying this collection of devices, our goal is to understand the effect of the phononic bandgap, electrodes, and material properties on the mechanical quality factors of phononic crystal resonators.  

We fabricate devices from thin-film, MgO-doped lithium niobate on silicon, where an etch process is used to define the phononic crystals in the LN. Subsequent lithographic masks add the aluminum electrodes and contact pads, after which the resonators are suspended by isotropically etching the silicon substrate. For material comparison, we also fabricate nominally identical devices from congruent-LN, which is known to have crystalline defects due to missing lithium atoms. A detailed fabrication process is presented in the supplemental information, with Fig.~1(b-d) showing the final fabricated resonators. Owing to their compact and highly parallelizable nature, we are able to fabricate 40 resonators per device in a $\sim50\times 200\,\mu\text{m}^2$ footprint. In order to better understand the effect of the aluminum electrodes on resonator coherence times, we design and fabricate resonators with two different electrode configurations: on-defect (Fig.~\ref{fig_device}(e)) and off-defect (Fig.~\ref{fig_device}(f)). The close proximity of the on-defect electrodes allows for over an order of magnitude enhancement in electromechanical coupling compared to the off-defect designs, coming at the cost of additional mechanical losses from the aluminum.

Inserting these devices at the end of a $50\,\Omega$ microwave transmission line allows for extraction of device properties via microwave reflection measurements, with representative results shown in Fig.~\ref{fig_device}(g). We fit each resonator's scattering parameter $S_{11}$ using the diameter correction method~\cite{Khalil2012}
\begin{align} \label{eq_S11}
    S_{11}(\omega) = 1 - e^{i\phi}\frac{Q}{\Qe }\frac{2}{1 + 2iQ\frac{\omega-\omega_r}{\omega_r}}\,,
\end{align}
where $\omega_r$ is the resonant frequency and the total quality factor $Q = (\Qe^{-1} + \Qi^{-1})^{-1}$ is comprised of the coupling $\Qe$ and internal $\Qi$ quality factors. Here $e^{i\phi}$ accounts for any circuit asymmetry caused by an impedance mismatch in the resonator's microwave environment, and avoids overestimating $\Qi$ for large asymmetries~\cite{Khalil2012,McRae2020}. All microwave measurements are performed at millikelvin temperatures in a dilution refrigerator, with standard shielding, thermalization, and filtering for superconducting resonators~\cite{McRae2020}.

Upon cooling the 40-resonator devices, we initially search for signatures of the phononic bandgap by classifying and characterizing the designed modes in the frequency range spanning $1.5-3.0\,\text{GHz}$ (see supplemental information for details). All resonators have the same mirror cell design with the simulated band structure shown in Fig.~2(a), while the defect size is swept to place resonant frequencies both in and out of the full phononic bandgap. In Fig.~2(b), measured internal quality factors $\Qi$ show over an order of magnitude enhancement for resonators inside the simulated bandgap relative to those outside. These measurements are performed at $800\,\text{mK}$ with high microwave power ($5\times10^6$ intra-cavity phonons) in order to maximize $\Qi$ and saturate any two-level systems loss channels. Despite being outside of the bandgap, modes still can exist in the $\sim2.1-2.8\,\text{GHz}$ range because the corresponding mirror cell band has opposite symmetry to the defect's eigenmode; the magnitude of this symmetry-dependent protection is likely limited by any physical asymmetry in the fabricated device~\cite{Patel2017a}. Above $2.8\,\text{GHz}$, the phononic band structure is both denser and contains bands of the same symmetry as the defect mode, causing these higher frequency modes to have significant leakage.

\begin{figure}[t]
    \centering
    \includegraphics[width=85mm]{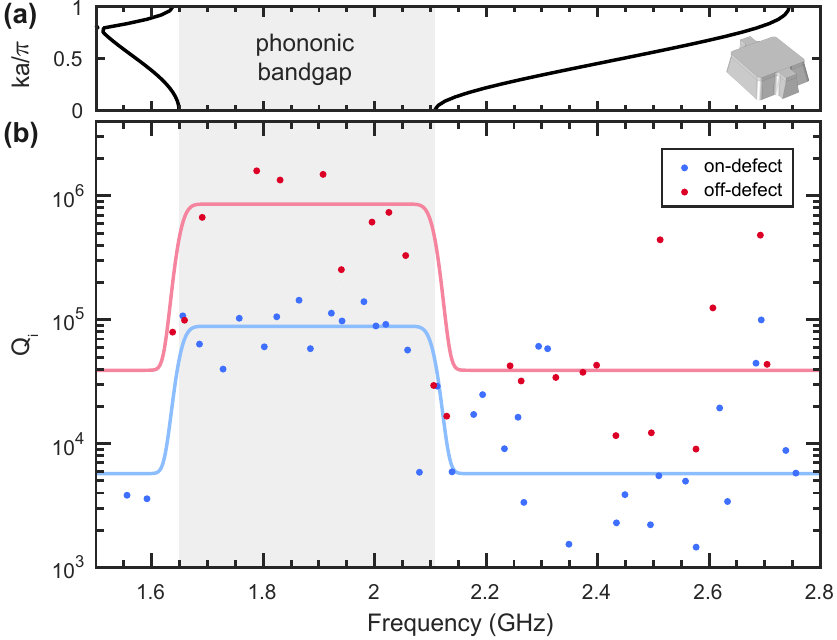}
    \caption{{\bf Signatures of the phononic bandgap.} (a)~Simulated band diagram for the mirror unit cell (pictured, right) forming the tethers of the defect site, with the full phononic bandgap from $1.65-2.1\,\text{GHz}$ shaded (grey). (b)~Internal quality factor $\Qi$ as a function of resonant frequency for the on-defect (blue) and off-defect (red) 40-resonator MgO-doped devices at $800\,\text{mK}$. By design, the resonant frequencies are swept across the simulated location of the phononic bandgap. The solid lines are guides to the eye, connecting the average $\Qi$ inside the simulated bandgap to the median $\Qi$ outside of the bandgap and demonstrating over an order of magnitude $\Qi$ enhancement in the protected region. The skewed distribution in $\Qi$ outside the bandgap makes the median a better indicator of typical device performance. }
    \label{fig_bandgap}
\end{figure}

After determining the location of the phononic bandgap, we explore the properties of these high-$\Qi$ modes for integration with superconducting quantum devices. Since the phononic bandgap eliminates all scattering losses, the $\Qi$ of these devices must be limited by other physical processes. We are primarily interested in how these devices operate when cooled to the millikelvin temperatures where operation in the quantum regime is possible. At such low temperatures, nonlinear processes such as Akhiezer, Landau-Rumer, and thermoelastic damping are greatly diminished~\cite{Liekens1971}, leading us to consider the resonators' two primary loss channels: (i)~two-level system (TLS) defects coupled to the mechanical resonance, and (ii)~ohmic and quasiparticle losses due to currents induced in the electrode metal via the piezoelectric effect. 

TLS defects are typically associated with two nearly-degenerate low-energy configurations of an amorphous material~\cite{Phillips1987}, and can couple to the strain or electric fields generated by excitation of a piezoelectric resonator. At single-phonon occupancy levels and millikelvin temperatures, TLS are no longer saturated, allowing them to cause dissipation and decoherence in coupled devices. Because these material losses are thought to dominate qubit and resonator lifetimes, TLS have been studied extensively in superconducting resonators~\cite{Gao2008,McRae2020}, as well as low- (${\scriptstyle\lesssim}\,100\,\text{MHz}$)~\cite{Hoehne2010,Riviere2011,Suh2013,Hauer2018} and high- (${\scriptstyle\gtrsim}\,1\,\text{GHz}$)~\cite{Manenti2016,Hamoumi2018,MacCabe2019,Andersson2020} frequency mechanical resonators.

Generically, TLS affect the resonator via resonant and relaxation processes, with the latter being strongly dependent on the dimensionality $d$ of the system~\cite{Behunin2016,Hauer2018,MacCabe2019}. In resonant processes, the TLS serve as an acoustic bath which the resonator can decay into, resulting in a temperature- and power-dependent contribution to the resonator loss~\cite{Phillips1987,Gao2008,Behunin2016}
\begin{align} \label{eq_Qi_res}
    \Qires^{-1} = F\deltaTLS \frac{\tanh \left(\frac{\hbar\omega_r}{2k_B T}\right)}{\sqrt{1 + \langle n \rangle / n_c}}\,.
\end{align}
Here, $\deltaTLS$ is the intrinsic TLS loss tangent at zero temperature, $F$ is the filling fraction of TLS in the host material, $\langle n \rangle$ is the average phonon number in the resonator, and $n_c$ is the critical phonon number for TLS saturation. Resonant TLS absorption does not depend on the system dimensionality -- up to a small numerical correction~\cite{Behunin2016} -- and so the reduction in $\Qires$ at sufficiently low temperature should be the same in both superconducting and $d$-dimensional mechanical resonators.

In contrast, relaxation absorption occurs due an interaction with off-resonant TLS, in which the oscillatory motion of the resonator causes local strain and electric field variations that perturb the TLS environment and modulate their energy levels. This modulation periodically displaces the TLS from  thermal equilibrium, with their subsequent relaxation back to equilibrium effectively damping the resonator. This process is strongly dependent on the dimensionality $d$ of the system, set by the device geometry, mode dispersion, and phonon bath density of states seen by thermally active TLS with frequencies $\omega_{\text{TLS}} \simeq k_B T/\hbar$. When limited by relaxation processes, the effective system dimension can be probed via the temperature dependence of the relaxation loss channel~\cite{Behunin2016,Hauer2018,MacCabe2019}, $\Qirel^{-1} \sim T^d$ (see supplemental information for further discussion).

\begin{figure}[t]
    \centering
    \includegraphics[width=85mm]{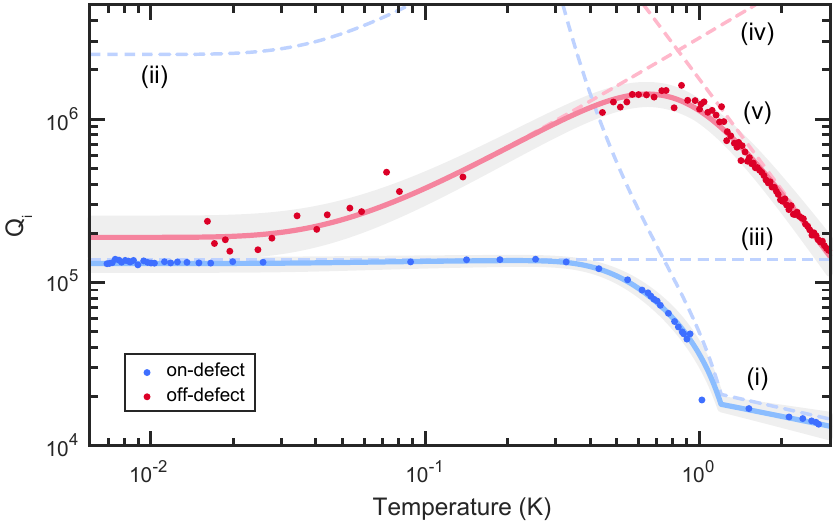}
    \caption{{\bf Quality factor measurements.} Internal quality factor $\Qi$ versus fridge temperature for an on-defect (blue) and an off-defect (red) MgO-doped device with resonant frequency $\omega_r/2\pi \simeq 2.0\,\text{GHz}$. The on-defect device is fit to (i) Mattis-Bardeen theory with (ii) resonant TLS decay at low temperatures and (iii) temperature-independent mechanical loss from the aluminum. Due to the absence of superconducting electrodes for the off-defect device, we fit to a TLS model which includes both (iv) resonant decay and (v) relaxation damping processes. Solid lines denote fits to each device's total model, with grey shaded regions corresponding to 95\% confidence intervals for the fits. We remark that on- and off-defect measurements were performed at high power, $\langle n \rangle  \simeq 5\times 10^6$ and $1.3\times10^4$ respectively, leading to a large power-enhancement in $\Qires$ at low temperature.}
    \label{fig_Qi_vs_T}
\end{figure}

\begin{figure}[t]
    \centering
    \includegraphics[width=85mm]{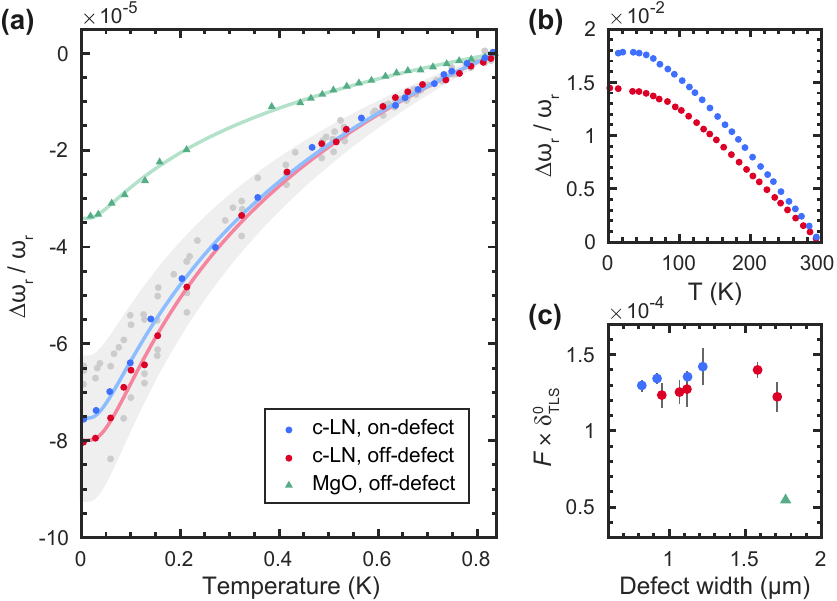}
    \caption{{\bf Anomalous low-temperature frequency shift.} (a)~Fractional shift in resonance frequency $\Delta \omega_r / \omega_r$ versus temperature, referenced to the resonator frequency at $800\,\text{mK}$. Devices fabricated from congruent-LN ($\bullet$) show dramatically different behavior than devices made from MgO-doped LN ($\blacktriangle$). Solid lines are fits to a TLS model (Eq.~\ref{eq_TLS_f_shift}) with a small incident thermal bath. For clarity, data from additional c-LN devices are shown as grey circles, and the shaded grey region indicates 95\% bounds on the c-LN devices' TLS loss. (b)~Fractional frequency shift referenced to room temperature. Resonators experience a large blueshift upon cooling, with the small redshift of (a) only occurring once off-resonant TLS are no longer thermally populated. (c) Total TLS loss $F\deltaTLS$ versus resonator defect width, with error bars indicating one standard deviation.}
    \label{fig_TLS}
\end{figure}

In order to identify loss channels in the phononic crystal resonators, we extract the internal loss $\Qi$ from microwave reflection measurements at a variety of fridge temperatures (Fig.~\ref{fig_Qi_vs_T}). For on-defect designs, we find that normal-metal losses in the aluminum dominate until the onset of superconductivity at $T_c \simeq 1.2\,\text{K}$. These normal-metal losses can be explained by the variation in electric potential on the surface of the defect site, where excitation of the fundamental shear mode piezoelectrically induces a gradient in the potential {\it within} each electrode. The resulting internal currents would then cause ohmic losses associated with the excitation of the resonance, not simply insertion loss from the leads. At intermediate temperatures below $T_c$, $\Qi$ is limited by thermal population of quasiparticles in the electrodes, described by the Mattis-Bardeen formulas for AC conductivity of a BCS superconductor~\cite{Mattis1958,Gao2008thesis,Zmuidzinas2012}.
With further reduction in temperature, resonant TLS losses outweigh superconducting loss channels, as evidenced by the absence of the frequency blueshift predicted by Mattis-Bardeen theory in Fig.~\ref{fig_TLS}(a). However, we suspect the grain-boundary mechanical losses of the polycrystalline aluminum electrodes~\cite{Ke1947} impose an approximately temperature-independent limit on $\Qi$, diminishing the visibility of resonant TLS loss $\Qires^{-1}(T)$ for $T \lesssim \hbar\omega_r/2 k_B$ in Fig.~\ref{fig_Qi_vs_T}.

For off-defect designs, we see markedly different behavior due to the off-defect resonator's electrodes being placed far from the mechanically active region. This means that aluminum's normal-metal, quasiparticle, and mechanical loss channels should be negligible for all temperatures, in agreement with the absence of any feature in $\Qi(T)$ near the $T_c$ of aluminum and the two orders of magnitude higher $\Qi(T_c)$ than on-defect devices. Instead, we find TLS relaxation absorption dominates for temperatures above $T\gtrsim 1\,\text{K}$, where $\Qirel^{-1} \sim T^d$ scales as the effective phonon bath dimension, $d=2.28\pm0.06$. Below $T\simeq300\,\text{mK}$, we estimate the wavelength of thermal phonons $\lambda_\text{th}\simeq1\,\mu\text{m}$ should be larger than both the LN thickness ($250\,\text{nm}$) and most feature sizes of the device, leading to a further reduction in the effective dimension. However, resonant TLS already contribute significantly to the device's $\Qi$ by this temperature, preventing observation of this geometry- and temperature-dependent damping for relaxation TLS in a mesoscopic device~\cite{Behunin2016}. Detailed analysis of both the on- and off-defect models for $\Qi(T)$ can be found in the supplemental information.

While sweeping the temperature, we also look for signatures of TLS by monitoring the resonant frequency of the mechanical mode (Fig.~\ref{fig_TLS}(a)). Due to interactions with resonant TLS, the relative frequency shift $\Delta\omega_r/\omega_r$ of the resonator is described by~\cite{Phillips1987,Gao2008}
\begin{align}\label{eq_TLS_f_shift}
    \!\!\!\!\!\frac{\Delta\omega_r}{\omega_r} = \frac{F\deltaTLS}{\pi}\! \left[ \text{Re}\!\left\{\!\Psi\!\left(\frac{1}{2}\!+\!\frac{\hbar\omega_r}{2\pi i k_B T}\right)\!\right\} \!-\!\ln\frac{\hbar\omega_{r}}{2\pi k_B T}\right]\!,\!\!\!\!\!
\end{align}
where $\Psi$ is the complex digamma function. This anomalous frequency redshift with decreasing temperature stands in contrast to the blueshifts associated with stiffening of the material (Fig.~\ref{fig_TLS}(b)) or the suppression of thermal quasiparticles in Mattis-Bardeen theory. Fits of $\Delta\omega_r/\omega_r$ in Fig.~\ref{fig_TLS}(a) to Eq.~\ref{eq_TLS_f_shift} provide an indirect method to extract the average total TLS loss $F\deltaTLS$ with high microwave probe power. In this measurement, far-detuned TLS still contribute significantly to the resonator's frequency shift at low temperature, thereby giving a spectrally-averaged value of TLS loss \cite{Pappas2011}. For both on- and off-defect devices made from congruent-LN, we report an average total TLS loss $F\deltaTLS = (1.28\pm0.09)\times10^{-4}$, while devices on MgO-doped LN achieve lower loss, $F\deltaTLS = (5.5\pm0.1)\times10^{-5}$. We note that the congruent-LN measurements include resonances both in and out of the phononic bandgap, and are consistent across several fabrication batches and cooldowns. In Fig.~\ref{fig_TLS}(c), we plot $F\deltaTLS$ versus the defect width to determine if TLS loss scales with geometry, as was found for coplanar waveguide resonators~\cite{Gao2008}. Both due to the narrow range of defect sizes, and ambiguity in the relative contributions of electric- and strain-coupled TLS to the filling fraction $F$ in a piezoelectric material, we are unable to differentiate between a bulk or surface TLS distribution. However, the significant decrease in TLS loss for MgO-doped LN suggests that the TLS loss may be mitigated through better material choices. For example, LN is known to have crystal defects due to missing lithium atoms, and these vacancies can be filled via re-lithiation or MgO-doping.

In summary, we have demonstrated phononic crystal resonators with internal quality factors $\Qi \simeq 10^5 - 10^6$ at high drive powers, with $5\times10^6$ phonons in the resonator. Detection of the fundamental phononic bandgap from $1.65-2.1\,\text{GHz}$ via internal quality factor $\Qi$ shows over an order of magnitude enhancement in $\Qi$ for modes inside the phononic bandgap. Through measurement of $\Qi$ versus temperature, we are able to demonstrate how normal-metal, quasiparticle, and TLS loss channels affect device performance. The best observed average total TLS loss $F\deltaTLS=(5.5\pm0.1)\times10^{-5}$ for devices made from MgO-doped lithium niobate implies a single phonon $\Qi\simeq 2\times 10^4$ is achievable with current generation devices. Understanding the origin of TLS in lithium niobate remains an outstanding challenge, requiring further advances in materials and fabrication.

\section*{Supplementary Material}
See \href{}{supplementary material} for information on device design and fabrication, resonator modeling and loss analysis, and applications to mass sensing.

\section*{Acknowledgments}
The authors would like to thank Cyndia Yu, Dale Li, Kevin K. S. Multani, Prof. Kent D. Irwin, and Prof. Martin Fejer for experimental support and helpful discussions. We acknowledge the support of the David and Lucille Packard Fellowship, and the Stanford University Terman Fellowship. This work was funded by the U.S. government through the Office of Naval Research (ONR) under grant No.~N00014-20-1-2422, the U.S. Department of Energy through Grant No. DE-SC0019174, and the National Science Foundation CAREER award No.~ECCS-1941826. E.A.W. was supported by the Department of Defense through the National Defense \& Engineering Graduate Fellowship. Device fabrication was performed at the Stanford Nano Shared Facilities (SNSF), supported by the National Science Foundation under grant No.~ECCS-1542152, and the Stanford Nanofabrication Facility (SNF). The authors wish to thank NTT Research for their financial and technical support.

\section*{Data Availability}
The data that support the findings of this study are available from the corresponding author upon request.

\pagebreak
\clearpage

\onecolumngrid
\vspace{\columnsep}
\begin{center}
\textbf{\large Supplemental information for ``Loss channels affecting lithium niobate phononic crystal resonators at cryogenic temperature"}
\end{center}
\vspace{\columnsep}
\twocolumngrid

\renewcommand{\theequation}{S\arabic{equation}}
\renewcommand{\thefigure}{S\arabic{figure}}
\renewcommand{\thetable}{S\arabic{table}}

\setcounter{equation}{0}
\setcounter{figure}{0}
\setcounter{table}{0}

\section{Design and simulations}\label{SI_sect_design}

The devices investigated in this experiment are nominally identical to those found in Ref.~\cite{Arrangoiz-Arriola2019}, which contains details of the phononic bandgap design and defect site simulations. Both the on-defect and off-defect resonators have the same bandgap design, with the only difference being the placement of the electrodes near the defect site. 

\begin{figure}[H]
    \centering
    \includegraphics[width=85mm]{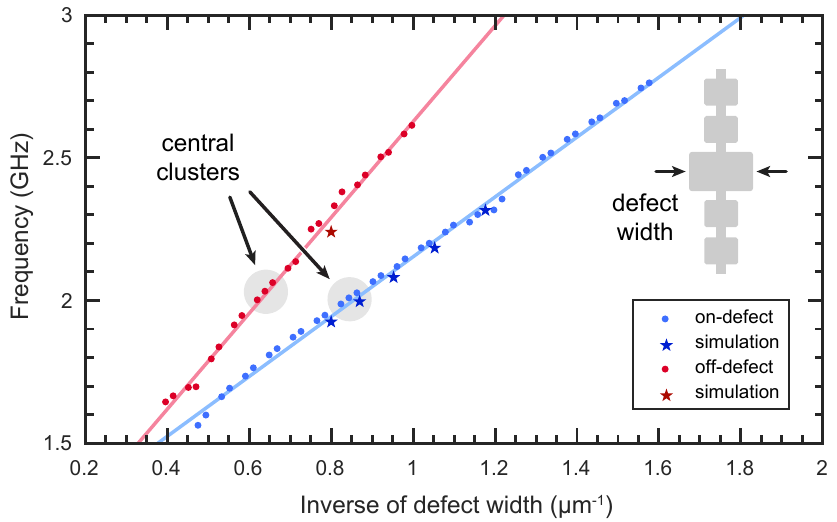}
    \caption{{\bf Frequency barcode.} Observed resonance frequencies versus the inverse of the defect width for both the on-defect (blue) and off-defect (red) 40-resonator devices. A linear fit shows the expected frequency scaling with defect size, also in agreement with finite-element simulations ($\star$). From the frequency barcode design, we expect resonances to lie on a grid with every third omitted; a cluster of three resonances in the barcode was placed roughly at the center of phononic bandgap. Although all 40 resonances are present for the on-defect device, the off-defect device has a substantial number of resonances outside of the full and symmetry-protected bandgap regions spanning $\sim1.5-2.8\,\text{GHz}$. Note that the defect width for each resonator is determined from SEMs.}
    \label{SI_fig_frequency_barcode}
\end{figure}

In order to correctly identify and distinguish the designed mechanical modes from any spurious modes on the 40-resonator device, a barcode pattern was embedded into the resonator frequency distribution. By positioning resonances on an equally-spaced $20\,\text{MHz}$ grid and then omitting every third, we can link each observed resonance frequency to its corresponding physical resonator. In Fig.~\ref{SI_fig_frequency_barcode}, the measured resonant frequency is plotted versus the inverse of defect width, showing both linear scaling and the barcode's pattern. The ability to classify resonances based on the frequency pattern and their coupling $\Qe$ is especially important outside of the bandgap, where the mode structure is not tightly controlled and undesired modes can emerge.

\section{Device fabrication}\label{SI_sect_fab}

\begin{figure}[H]
    \centering
    \includegraphics[width=85mm]{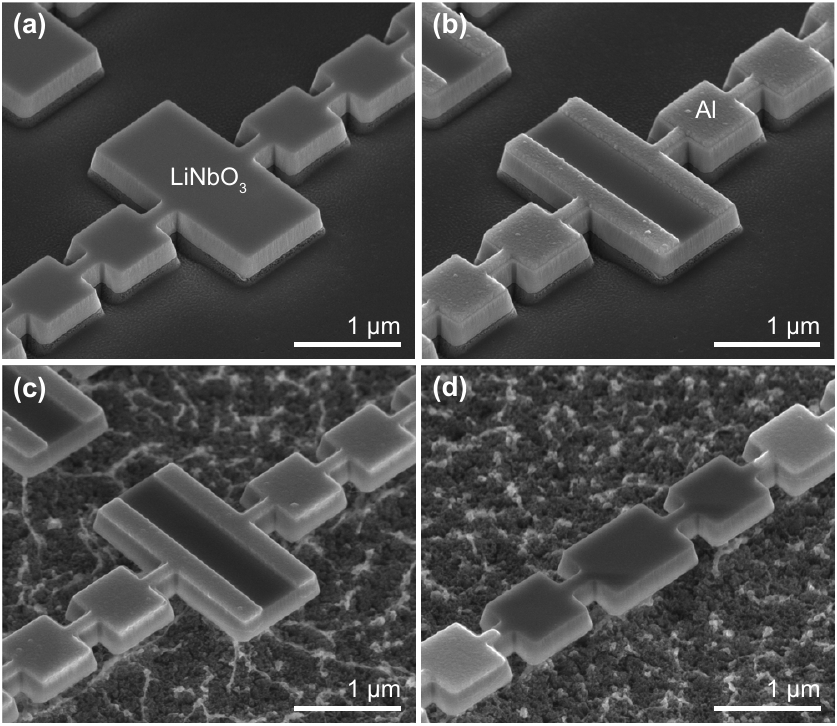}
    \caption{{\bf Device fabrication.} (a)~Patterned lithium niobate (LiNbO$_3$) mesa on silicon substrate after e-beam lithography, argon milling, and acid clean. (b)~Addition of aluminum electrodes via e-beam lithography and liftoff. (c,d)~Final on-defect and off-defect devices after release of structure using a XeF$_2$ dry etch.} 
    \label{SI_fig_fab_progression}
\end{figure}

We fabricate the mechanical resonators from a $500\,\text{nm}$ film of X-cut lithium niobate on a high-resistivity silicon substrate ($\rho > 3\,\text{k}\Omega\cdot\text{cm}$). In the first step, the lithium niobate is thinned to $250\,\text{nm}$ via blanket argon milling, after which electron-beam (e-beam) lithography is used to define the phononic crystal resonators using a negative resist (HSQ). The HSQ pattern is transferred to the lithium niobate through an angled argon milling step~\cite{Wang2014}. Subsequently, we strip the remaining resist and redeposited lithium niobate using an acid clean (Fig.~\ref{SI_fig_fab_progression}(a)), allowing us to proceed to the metal fabrication.

In the first metal layer, the device contact pads are defined via photolithography, electron-beam evaporation, and liftoff. Next, we fabricate the $50\,\text{nm}$ thick aluminum electrodes directly on top of the lithium niobate structures using e-beam lithography and liftoff (Fig.~\ref{SI_fig_fab_progression}(b)). The alignment of the electrodes to the existing phononic crystals is critical, and we typically obtain better than $\sim10\,\text{nm}$ alignment precision. In the final metal layer, we form superconducting connections between the electrodes on the silicon substrate and those on top of the $250\,\text{nm}$ thick lithium niobate. This bandaging process~\cite{Dunsworth2017} involves {\it in situ} milling of the electrode's native aluminum oxide, followed by electron-beam evaporation of $500\,\text{nm}$ of aluminum in order to climb the lithium niobate mesa.

The final step requires a masked release of the phononic crystal resonators (Fig.~\ref{SI_fig_fab_progression}(c,d)). After photolithographically defining the release area, we use a highly selective XeF$_2$ dry etch to undercut the lithium niobate structures from the silicon substrate~\cite{Vidal-alvarez2017}. The release mask is then stripped in solvents, and we package the devices for cryogenic measurement.

\section{Cryogenic Characterization}\label{SI_sect_cryo_measurements}

To characterize the mechanical resonators at cryogenic temperatures, we perform reflection measurements using a vector network analyzer (VNA, Rohde-Schwarz ZNB20). Our measurements from $T=300\,\text{K}$ to $4\,\text{K}$ are carried out in a $4\,\text{K}$ tabletop cryostat (Montana Nanoscale Workstation s200). Before cooling the sample, we first calibrate the VNA ports: by wirebonding to an on-chip open, short, and 50\,$\Omega$ match, we are able to de-embed the microwave cabling and wires up to the device bond pads. After packaging and calibration, the resonators are probed at various temperatures during the cooling of the cryostat. From each reflection measurement, we extract the internal and external quality factors, and numerically adjust for small, temperature-dependent gain and phase fluctuations in the calibration.

Measurements from $T=4\,\text{K}$ to $10\,\text{mK}$ are performed in a dilution refrigerator (Bluefors BF-LD250), with the sample surrounded by aluminum and cryoperm magnetic shielding. To ensure the device is well-thermalized, microwave lines leading to the device are filtered and attenuated at each temperature stage, and multiple isolators are used to route the reflected output signal to a high electron mobility transistor (HEMT) amplifier at $3\,\text{K}$. 

\section{Equivalent models of a mechanical resonator}\label{SI_sect_rosetta}

\begin{figure*}[p]
    \centering
    \includegraphics[width=170mm]{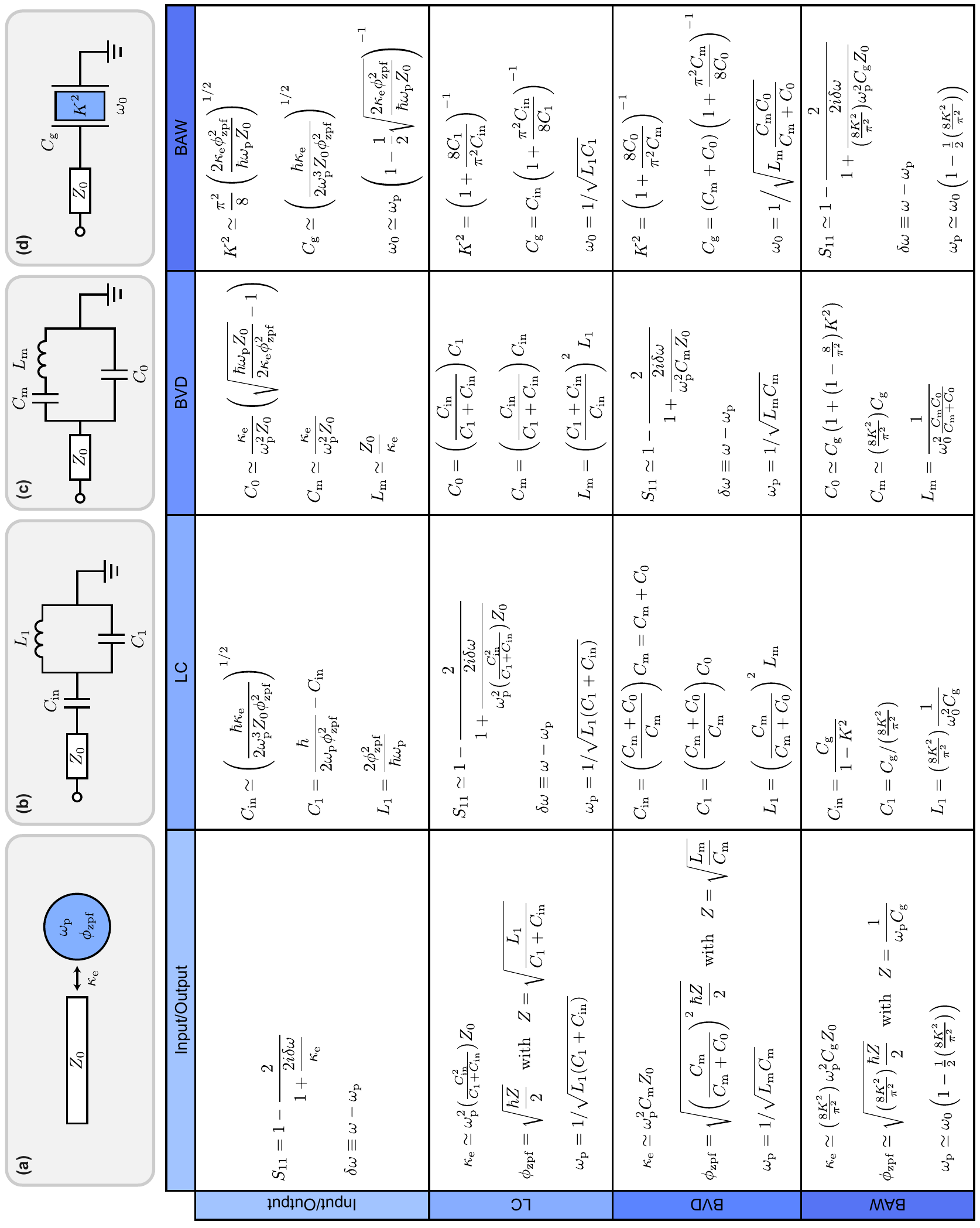}
    \caption{{\bf Rosetta stone.} A mechanical resonator can equivalently be described within the frameworks of (a)~input/output theory, (b)~a capacitively coupled $LC$ circuit, (c)~the Butterworth-van Dyke (BVD) circuit, or (d)~a thin-film bulk acoustic wave (BAW) resonator. This table provides conversion between the various model parameters (see Section~\ref{SI_sect_rosetta} for definitions of each variable). Here, an approximately equals sign ($\simeq$) indicates that the formula is only valid in a weak coupling approximation, e.g.~$\Ksq \ll 1$~or~$\omega\Cin\Z \ll 1$.}
    \label{SI_fig_rosetta}
\end{figure*}

The linear mechanical resonators found in this work can be described within the framework of input/output theory (I/O), a capacitively coupled $LC$ circuit ($LC$), the Butterworth-van Dyke circuit (BVD), or a thin-film bulk acoustic wave resonator (BAW). In an effort to facilitate comparison between these equivalent models, we present a conversion table in Fig.~\ref{SI_fig_rosetta}, while Table~\ref{SI_table_params} gives simulated device parameters within the context of each model. Below, we provide a brief synopsis of each framework and highlight some of their common applications.

In a quantum-optical input/output formalism~\cite{walls2007quantum}, a semi-infinite waveguide $\Z$ is coupled to a harmonic oscillator at rate $\kappae$ (Fig.~\ref{SI_fig_rosetta}(a)). Here, the harmonic oscillator is described by its resonant frequency $\omegap$ and the zero-point fluctuations of its position (flux) operator $x_\text{zpf}$ ($\phizpf$). The flux operator is commonly used in circuit quantum electrodynamics (cQED), where the node flux $\phi(t) = \int^t V(\tau) d\tau$ is used as the position coordinate of the oscillator. This description of the mechanical oscillators most naturally extends to the Hamiltonian-based formalisms of cQED, where phononic systems are currently being integrated with superconducting qubits in various architectures~\cite{OConnell2010,Arrangoiz-Arriola2016,Chu2017,Chu2018,Satzinger2018,Arrangoiz-Arriola2019,Sletten2019,Chu2020}.

\begin{table}[t]
    \centering
    \begin{tabular}{|c|c|c|c|}
    \hline \hline
    ~Model~ & ~Parameter~ & ~On-defect~ & ~Off-defect~ \\ \hline
    I/O & $\begin{aligned}
            &\kappae/2\pi \\
            &\phizpf \\
            &\omegap/2\pi \\
          \end{aligned}$
        & $\begin{aligned}
            85&\,\text{kHz} \\
            1.45&\,\text{fm} \\
            1.925&\,\text{GHz} \\
          \end{aligned}$
        & $\begin{aligned}
            0.75&\,\text{kHz} \\
            0.55&\,\text{fm} \\
            2.240&\,\text{GHz} \\
          \end{aligned}$
        \\ \hline
    $LC$& $\begin{aligned}
            &\Cin \\
            &\Cone \\
            &\Lone \\
          \end{aligned}$
        & $\begin{aligned}
            390&\,\text{aF} \\
            1.7&\,\text{fF} \\
            3.3&\,\mu\text{H} \\
          \end{aligned}$
        & $\begin{aligned}
            77&\,\text{aF} \\
            12&\,\text{fF} \\
            400&\,\text{nH} \\
          \end{aligned}$
        \\ \hline
    BVD & $\begin{aligned}
            &\Czero \\
            &\Cm \\
            &\Lm \\
          \end{aligned}$
        & $\begin{aligned}
            320&\,\text{aF} \\
            73&\,\text{aF} \\
            93&\,\mu\text{H} \\
          \end{aligned}$
        & $\begin{aligned}
            77&\,\text{aF} \\
            0.48&\,\text{aF} \\
            10.6&\,\text{mH} \\
          \end{aligned}$
        \\ \hline
    BAW & $\begin{aligned}
            &\Ksq \\
            &\Cg \\
            &\omegao/2\pi \\
          \end{aligned}$
        & $\begin{aligned}
            0.&22 \\
            300&\,\text{aF} \\
            2.136&\,\text{GHz} \\
          \end{aligned}$
        & $\begin{aligned}
            0.0&076 \\
            77&\,\text{aF} \\
            2.247&\,\text{GHz} \\
          \end{aligned}$
        \\ \hline \hline
    \end{tabular}
    \caption{Simulated device parameters for the on-defect and off-defect designs, both with the same defect site dimensions $1 \times 1.25\,\mu\text{m}^2$. We note the large difference in $\kappae$, $\Cin$, and $\Ksq$ between the two resonators due to the dramatically different electrode designs. This results in $\sim 10$ times higher $g$ for the on-defect design when coupling the mechanical system to another oscillator (Section~\ref{SI_sect_g_rosetta}).}
    \label{SI_table_params}
\end{table}

The $LC$ circuit found in Fig.~\ref{SI_fig_rosetta}(b) consists of a parallel $\Lone\Cone$ capacitively coupled via $\Cin$ to a transmission line of characteristic impedance $\Z$. The admittance function takes the form

\begin{align}
    Y_{LC} = i\omega \Cin \;\frac{1-\omega^2\Lone\Cone}{1 - \omega^2\Lone(\Cone + \Cin)}\;,
\end{align}
from which we find a pole in the admittance at $\omegap = 1/\sqrt{\Lone(\Cin + \Cone)}$, and a zero at $\omegao = 1/\sqrt{\Lone\Cone}$. Typically the input capacitance is weak, $\omega\Cin\Z\ll 1$, leading to the approximate equations found in Fig.~\ref{SI_fig_rosetta}. This model takes an equivalent circuit perspective on the mechanical system, drawing direct comparisons to e.g. the analysis of microwave resonators.

Due to ambiguity in one's choice of Foster synthesis, the BVD circuit~\cite{IEEE88,Larson2000} in Fig.~\ref{SI_fig_rosetta}(c) is entirely equivalent to the $LC$ model. Here, a transmission line $\Z$ is directly connected to a series $\Lm\Cm$ in parallel with a static capacitance $\Czero$, leading to an admittance function    
\begin{align}\label{SI_eq_BVD}
    Y_\text{BVD} = i\omega (\Czero + \Cm) \;\frac{1-\omega^2\Lm\left(\frac{\Cm\Czero}{\Cm + \Czero}\right)}{1 - \omega^2\Lm\Cm}\;.
\end{align}
Similar to the $LC$ model, the admittance has a pole at $\omegap$, a zero at $\omegao$, and a electrostatic capacitance as $\omega \rightarrow 0$. The BVD model is commonly used to analyze mechanical resonators, with the advantage that it separates out the static capacitance $C_0$ of the electrodes from the mechanically active portion $\Lm\Cm$.

\begin{figure*}[t!]
    \centering
    \includegraphics[width=170mm]{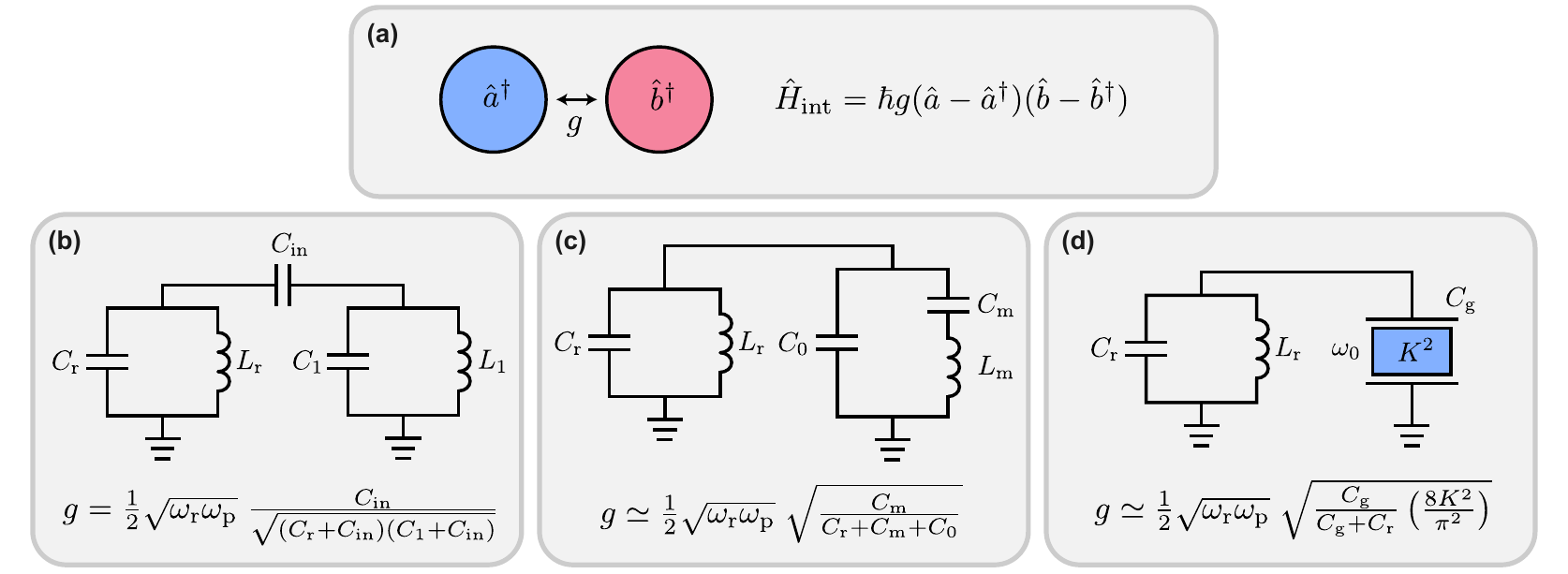}
    \caption{{\bf Resonator-resonator coupling.} (a) In a quantum optics picture, the interaction $\hat{H}_\text{int}$ between oscillators $\hat{a}^\dagger$, $\hat{b}^\dagger$ is governed by their coupling rate $g$, which sets the strength of the interaction. When designing the interaction between mechanical and electrical resonators, it is often advantageous to work in this picture, where $g$ depends on the details of the equivalent circuits. (b, c, d) Coupling rate $g$ for the interaction between a $LC$ circuit and the $LC$, BVD, and BAW models of a mechanical system, respectively. Here, $\omegar$ and $\omegap$ correspond to each resonator's pole in admittance $Y(\omega)$, and are capacitively loaded by the presence of the other resonator (see Eq.~\ref{SI_eq_omega1}). An approximately equals sign ($\simeq$) indicates that the formula is only valid in a weak coupling approximation, e.g.~$\Ksq \ll 1$ or $\Cin \ll \Cone,C_2$.}
    \label{SI_fig_g_rosetta}
\end{figure*}

The final model we consider is that of a thin-film bulk acoustic wave resonator, in which a thin-film piezo is driven by a gate capacitance $C_g$ (Fig.~\ref{SI_fig_rosetta}(d)). The piezo is characterized by its effective electro-acoustic coupling constant $\Ksq$, its film thickness $b$, and the speed of sound $v$ in the crystal. Ignoring edge effects, the BAW model yields an admittance of the form~\cite{Cleland2004,Hashimoto2009}
\begin{align}
    Y_\text{BAW} = i\omega \Cg \left[ 1 - \Ksq \frac{\tan \left(\omega b/2v\right)}{\omega b/2v}\right]^{-1}\,,
\end{align}
relating the induced electrode current to the voltage between them. The first zero in the admittance function occurs at $\omegao = 2\pi v/2b$, corresponding to the fundamental excitation of the film, while the first pole occurs at $\omegap < \omegao$. Although phononic crystal devices are not bulk acoustic wave resonators, this model illuminates how the piezo's material properties should affect device performance, and avoids abstracting the mechanical nature of the resonator via an equivalent circuit. 

Notably, the admittance functions for the $LC$, BVD, and BAW models all have the same generic behavior, with the only difference being the variables used to describe the system. In each admittance, the first term gives the electrostatic behavior of the system as $\omega \rightarrow 0$, while the second term gives the zeros $\omegao$ and poles $\omegap < \omegao$ in admittance that characterize the strength of the electro-acoustic coupling $\sim (\omegao - \omegap$). Each of these models provide the designer different insights for the design, characterization, and applications of mechanical systems. On the other hand, the input-output model readily extends to a quantized Hamiltonian of creation/annihilation operators, allowing for integration with quantum circuits. Consequently, each model has strengths and weaknesses depending on the level of abstraction of the mechanical system, whether the mechanical resonators are part of a larger circuit, and the device's final application.

\section{Resonator-resonator coupling} \label{SI_sect_g_rosetta}

Coupling mechanical resonators with electrical circuits is an integral aspect of the study of electro-mechanical systems. The coupling between the two resonators can be described by their interaction strength $g$, which is set by the details of each system's equivalent circuit. Depending on whether the mechanical resonators are modeled in the $LC$, BVD, or BAW pictures, $g$ takes the various forms found in Fig.~\ref{SI_fig_g_rosetta} when the mechanical system is coupled to a $L_{\text{r}}C_{\text{r}}$ resonator with resonant frequency $\omegar$.

Across all interaction models, the coupling rate depends on the geometric mean of the two resonant frequencies $\omegar$, $\omegap$ corresponding with each resonator's pole in admittance. We note that both $\omegar$ and $\omegap$ are loaded by the presence of the other coupled oscillator, e.g.
\begin{equation} \label{SI_eq_omega1}
     \omegar = \frac{1}{\sqrt{L_{\text{r}} \left(C_{\text{r}} + \frac{\Cone\Cin}{\Cone + \Cin}\right)}} \simeq \frac{1}{\sqrt{L_{\text{r}} \left(C_{\text{r}} + \Cin\right)}}\;,
\end{equation}
where the approximation holds in the case of weak coupling, $\Cin\ll\Cone,C_{\text{r}}$.

For our device, we can use finite-element simulations to predict the relative coupling strength of the on-defect and off-defect designs using the results of Table~\ref{SI_table_params}. Without needing to know the details of the coupled $LC$ resonator, we estimate the ratio $g_\text{on}/g_\text{off} \approx 12$ for mechanical resonators with identical dimensions. 

\section{Qubit-Resonator Coupling}\label{SI_sect_qubit_coupling}

\begin{figure}[t]
    \centering
    \includegraphics[width=85mm]{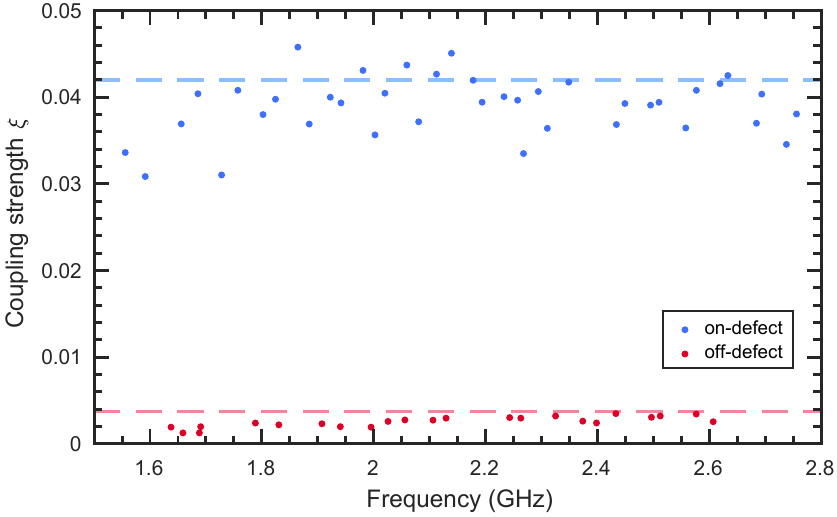}
    \caption{{\bf Coupling strength.} The dimensionless coupling strength $\xi$ is extracted from microwave reflection measurements $S_{11}$ according to Eq.~\ref{SI_eq_xi} for both the on-defect (blue) and off-defect (red) devices. Finite-element simulations of the resonators (dashed lines) closely match the measured values.}
    \label{SI_fig_qubit_coupling}
\end{figure}

When integrating phononic crystal resonators with superconducting quantum circuits, it is desirable to achieve strong coupling between the two systems. Replacing one of the resonators in the previous section with a transmon qubit yields a coupling rate~\cite{Arrangoiz-Arriola2016}
\begin{equation} \label{SI_eq_g_qubit}
    g = \xi \sqrt{\alpha (\omega_{\text{ge}} + \alpha)}
\end{equation}
which depends on the qubit ground to excited state transition frequency $\omega_{\text{ge}}$ and anharmonicity $\alpha$. As derived in Ref.~\cite{Arrangoiz-Arriola2019}, $\xi$ is a dimensionless coupling-strength parameter entirely determined by the details of the mechanical system,
\begin{equation}
    \xi = \left(\frac{\Cin}{\Cone + \Cin}\right)\left(\frac{\hbar^2}{4e^4}\frac{\Cone + \Cin}{\Lone}\right)^{1/4}\;.
\end{equation}
Although $\xi$ depends on all three parameters of the $LC$ model, we can switch to the input/output model where notably it does not depend on the zero-point fluctuations $\phizpf$ of the resonator,
\begin{equation} \label{SI_eq_xi}
    \xi = \sqrt{\frac{\hbar}{2e^2}\frac{\kappae}{Z_0\omegap}}\;.
\end{equation}
Since $\kappae$ and $\omegap$ are easily extracted from scattering parameter measurements, we can measure the expected coupling strength without the need for more complex device characterization or full integration with the qubit system. In Fig.~\ref{SI_fig_qubit_coupling}, we show the implied coupling strength $\xi$ from microwave reflection measurements for on- and off-defect devices. The average $\xi\simeq3.9\times10^{-2}$ for on-defect modes in the bandgap yields a coupling rate $g/2\pi\simeq 23\,\text{MHz}$, assuming realistic transmon parameters $\omega_{\text{ge}}/2\pi = 2.4\,\text{GHz}$ and $\alpha/2\pi = 140\,\text{MHz}$. This coupling is roughly an order of magnitude higher than the $\xi\simeq2.5\times10^{-3}$ for the off-defect modes, meaning they would need a correspondingly higher $\Qi$ to preserve the number of quantum gate operations $n_g \sim g \Qi/\omegap$ possible before energy relaxation of the mechanical system.

\section{Modeling resonator losses}
In the following sections we describe the methods used to model resonator loss and obtain the temperature dependence of $\Qi(T)$ for the fits in Fig.~\ref{fig_Qi_vs_T} in the main text.
\subsection{Modified Butterworth-van Dyke model}\label{SI_sect_mBVD}
\begin{figure}[t]
    \centering
    \includegraphics[width=85mm]{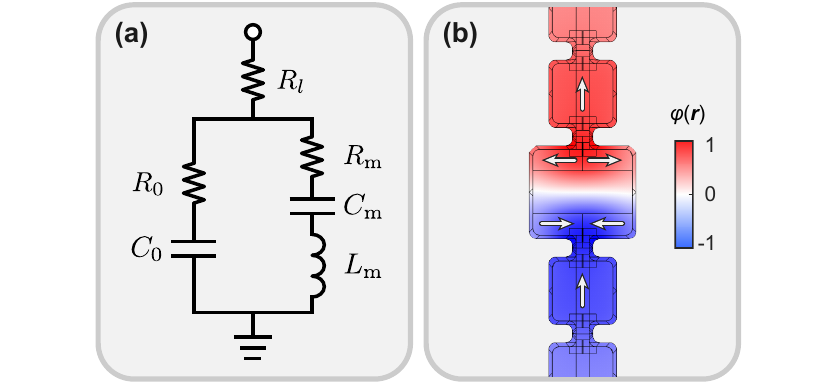}
    \caption{{\bf Modeling resonator losses.} (a)~ Modified Butterworth-van Dyke equivalent circuit model for a mechanical resonator. The addition of resistors $R_l$, $\Rm$, and $R_0$ describe lead, mechanical, and dielectric losses respectively. (b)~Finite-element simulations of the electrostatic potential $\varphi(\bf{r})$ associated with the strongly-coupled mode in an on-defect design. Arrows indicate the direction of decreasing potential, showing how the mode's mechanical deformation induces an electric potential gradient within each electrode.}
    \label{SI_fig_mBVD}
\end{figure}

In order to model the various loss channels of the resonator, we use the modified Butterworth-van Dyke (mBVD)\cite{Larson2000} equivalent circuit shown in Fig.~\ref{SI_fig_mBVD}(a) by adding resistors to the BVD model. Here, the resistor $\Rm$ is associated with mechanical losses, $R_0$ describes dielectric losses, and $R_l$ gives the resistance of the electrical leads. For the devices considered in this work, $R_l$ only contributes to the total insertion loss, so we can safely take $R_l = 0$. Solving for the admittance of the circuit in Fig.~\ref{SI_fig_mBVD} yields
\begin{align}
    Y_\text{mBVD} = i\omega (\Czero + \Cm) \;\frac{1-\frac{\omega^2}{\omegaot^2} + i\frac{1}{Q_{i,0}}\frac{\omega}{\omegaot}}{1-\frac{\omega^2}{\omegapt^2} + i\frac{1}{Q_{i,\text{p}}}\frac{\omega}{\omegapt}}\;,
\end{align}
where frequency $\omegaot$ ($\omegapt$) and internal quality factor $Q_{i,0}$ ($Q_{i,\text{p}}$) are associated with the zero (pole) in admittance. Note that $\omegaot$ and $\omegapt$ are loaded relative to the lossless case in Eq.~\ref{SI_eq_BVD}, with the pole now occurring at
\begin{align}
    \frac{1}{\omegapt^2} = \Lm\Cm \left(1 + \frac{\Rm R_0\Czero}{\Lm}\right)\;.
\end{align}
We are primarily concerned with the internal quality factor of the pole $\omegapt$ in admittance,
\begin{align} \label{SI_eq_Qip}
    Q_{i,\text{p}} \simeq \frac{1}{\omegapt\Rm\Cm}\;.
\end{align}
In the following sections, we take $\Lm$, $\Cm$, and $\Czero$ from simulations of the lossless case in Table~\ref{SI_table_params}.

\subsection{On-defect $\Qi(T)$ model}\label{SI_sect_on_defect_model}
For on-defect devices, the placement of aluminum on the mechanically active region induces normal-metal, quasiparticle, and mechanical losses not seen in the off-defect designs. In order to obtain the fit of $\Qi(T)$ in Fig.~\ref{fig_Qi_vs_T} of the main text, we use the mBVD model described above in combination with the Mattis-Bardeen theory for AC conductivity of a BCS superconductor~\cite{Mattis1958,Gao2008thesis,Zmuidzinas2012}.

Due to the piezoelectricity of LN, when the defect site is deformed it creates an electric potential which varies across the surface of the resonator, see Fig.~\ref{SI_fig_mBVD}(b). Because the potential varies within each electrode, this creates internal currents and results in ohmic losses from the temperature dependent electrode resistivity $\rho(T)$. These losses are not associated with the leads ($R_l$ in the mBVD model) because they are being driven by the electromechanical mode itself, leading us to lump them into $\Rm$ for the mechanical motion. To model the normal-metal $R_n$ contribution to $\Rm$ for temperatures above the $T_c$ of aluminum, we assume
\begin{align}
    R_n(T) &= g_{e}\rho(T) \\
    &= g_e\rho_n \times (T/T_c)^\beta\;, \nonumber
\end{align}
where $\rho_n$ is the normal-state resistivity just above $T_c$, $\beta$ determines the resistivity's scaling with temperature, and $g_{e}$ is a geometrical factor determined by the electrode geometry and electric potential profile of the mode. 

For the on-defect resonator measured in Fig.~\ref{fig_Qi_vs_T}, we can estimate the expected $\Qi$ from normal-metal losses just above $T_c$ using finite-element simulations~\cite{comsol2013} of joule heating. Taking $\rho_n = 1.5-3\times10^{-8}\,\Omega\cdot\text{m}$ for thin-film aluminum\cite{Park1995}, we obtain a simulated $\Qi \simeq 1.2-2.4\times10^4$ and $g_{e}\simeq 3.0\,\text{nm}^{-1}$ for this geometry. This is in good agreement with the $\Qi\simeq2\times10^4$ measured in a variety of on-defect devices just above $T_c$.

For temperatures below $T_c$, $\Rm$ has a contribution from thermal quasiparticles in the superconducting electrodes. In the thin film limit, the surface impedance $Z_s$ of a superconductor with complex conductivity $\sigma(T) = \sigma_1(T) - i\sigma_2(T)$ is given by~\cite{Gao2008thesis}
\begin{align}\label{SI_eq_Zs}
    Z_s(T) = \frac{1}{(\sigma_1 - i\sigma_2)d}\;,
\end{align}
for a film of thickness $d$. Here $\sigma_1(T)$ and $\sigma_2(T)$ can be calculated by integrating the full Mattis-Bardeen equations for complex conductivity~\cite{Mattis1958,Gao2008thesis,Zmuidzinas2012}. From the surface impedance, we can determine the superconducting contribution $R_s$ to the mBVD effective resistance $\Rm$ as
\begin{align}
    R_s(T) &= g_e d \;\text{Re}(Z_s) \\
    &= g_e\frac{\sigma_1}{\sigma_1^2 + \sigma_2^2}\;. \nonumber
\end{align}
We note that as $T \rightarrow T_c$, the complex conductivity approaches the normal-state limit, $\sigma \rightarrow \sigma_n$, so that $R_s(T_c) = R_n(T_c)$.

For elastic losses, we observe that on-defect devices' $\Qi$ plateaus at intermediate temperatures $T \simeq 500\,\text{mK}$, a region in which off-defect devices begin to show strong signatures of resonant TLS loss. The notable difference in $\Qi$ and temperature-dependence between the two device styles leads us to believe that the on-defect devices may be limited by viscous losses of the grain-boundaries in the polycrystalline aluminum electrodes~\cite{Ke1947} beginning at intermediate temperatures. We model this elastic loss channel as a temperature-independent effective resistance $R_e$.

\begin{table}[t]
    \centering
    \begin{tabular}{|cc|cc|}
        \multicolumn{4}{l}{On-defect:} \\
        \hline \hline
        \multicolumn{2}{|c|}{Model constants} & \multicolumn{2}{c|}{Model fit parameters} \\ \hline
        ~$\omegapt/2\pi$~ & ~$1.998\,\text{GHz}$~ & ~$R_n(T_c)=g_e\rho_n$ & $52.8\pm 1.9\,\Omega$ \\
        $\Cm$ & $73\,\text{aF}$ & $R_e$ & $7.9 \pm 0.1\,\Omega$ \\
        $\Delta_0$ & $0.182\,\text{meV}$  & $\beta$ & $0.38 \pm 0.05$ \\
        $T_c$ & $1.2\,\text{K}$ & ~$\Qires(n)$~ & ~$(2.5 \pm 0.7) \times 10^6$~ \\
        $\langle n \rangle$ & $5\times 10^6$ & $n_c$ & $(2.6 \pm 1.6)\times 10^2$\\ \hline \hline
        \multicolumn{4}{l}{~} \\
        \multicolumn{4}{l}{Off-defect:} \\
        \hline \hline
        \multicolumn{2}{|c|}{~~~Model constants~~~} & \multicolumn{2}{c|}{~~~Model fit parameters~~~} \\ \hline
        $\omegap/2\pi$ & $1.910\,\text{GHz}$& $d$ & $2.28\pm.06$ \\
        $T_0$ & $800\,\text{mK}$ & $\Qirel(d)$ & $(3.0 \pm 0.1) \times 10^6$ \\
        $\langle n \rangle$ & $1.3 \times 10^4$ & $\Qires(n)$ & ~$(1.42 \pm 0.08) \times 10^5$~ \\
        & & $n_c$ & $(2.1 \pm 0.3)\times 10^2$\\ \hline \hline
    \end{tabular}
    \caption{{\bf Model parameters for $\Qi$ versus temperature.} See text for definitions of parameters and constants. Note that the critical phonon number $n_c$ is estimated from the fit value of $\Qires(n)$, not directly modeled.}
    \label{SI_table_Qi_T_params}
\end{table}

Combining all loss channels, we can model the temperature dependence of the mBVD effective resistance as
\begin{align} \label{SI_eq_Rm}
    \Rm(T) = R_e + 
    \begin{cases}
      R_n(T)\;, & T \ge T_c \\
      R_s(T)\;, & T < T_c\;.\\
    \end{cases}  
\end{align}
To incorporate resonant TLS losses (Eq.~\ref{eq_Qi_res}) into the model, we approximate the power-enhanced quality factor $\Qires^{-1}(n) \equiv F\deltaTLS (1 + \langle n \rangle / n_c)^{-1/2}$ as constant over the temperature range of interest, such that
\begin{align} \label{SI_eq_Qires}
    \Qires^{-1}(T) = \Qires^{-1}(n) \tanh \left(\frac{\hbar\omega_r}{2k_B T}\right) \;.
\end{align}
For the full fit to $\Qi(T)$ for the on-defect device in Fig.~\ref{fig_Qi_vs_T},
we combine Eqs.~\ref{SI_eq_Qip},~\ref{SI_eq_Rm},~and~\ref{SI_eq_Qires},
\begin{align}
    \Qi^{-1}(T) = \Qires^{-1}(T) + Q_{i,\text{p}}^{-1}(T)\;.
\end{align}
The best fit parameters and model constants are provided in Table~\ref{SI_table_Qi_T_params}. For model constants, we obtain $\omegapt$ and the average intra-cavity phonon number $\langle n \rangle$ from spectroscopic measurement, $\Cm$ from simulation (Table~\ref{SI_table_params}), and assume typical values for the superconducting energy gap $\Delta_0$ and critical temperature $T_c$ of aluminum. Using $\langle n \rangle$, the $F\deltaTLS$ from frequency redshift measurements in Fig.~\ref{fig_TLS}, and the fit $\Qires(n)$, we can estimate the critical phonon number $n_c$ for resonant TLS saturation. Notably the $\Qi(T)$ measurements for the on-defect device were taken in the high-power regime, so the power-enhanced $\Qires(n)$ is significantly higher than for the off-defect device.

\begin{figure}[t]
    \centering
    \includegraphics[width=85mm]{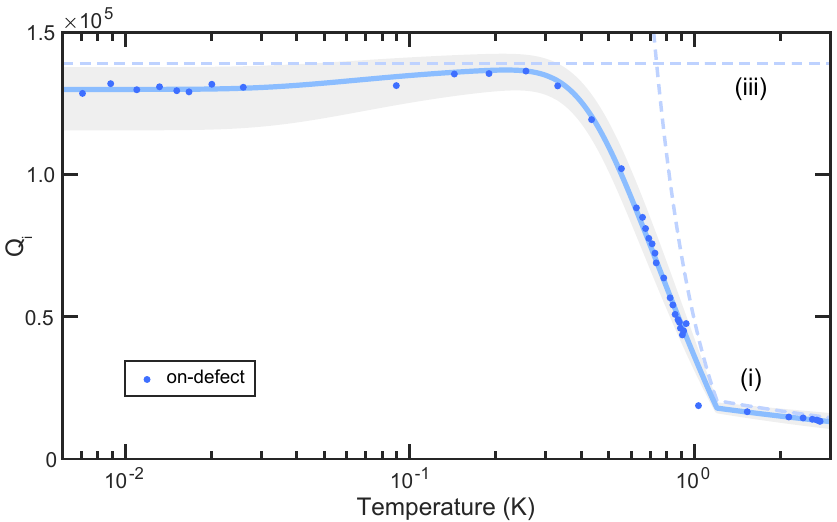}
    \caption{{\bf Quality factor measurements.} Internal quality factor $\Qi$ versus fridge temperature for an on-defect, MgO-doped device. The data is reprinted from Fig.~\ref{fig_Qi_vs_T} in the main text for visual clarity. The solid line denotes a fit to the total model, with grey shaded regions corresponding to 95\% confidence intervals. Loss channels from (i) Mattis-Bardeen theory and (iii) temperature-independent mechanical loss from the aluminum are shown as dashed lines. Resonant TLS decay (labeled (ii) in Fig.~\ref{fig_Qi_vs_T}) is still included in the model, but is not shown so that the behavior of the $\Qi$ data remains visible.}
    \label{SI_fig_Qi_vs_T_ondefect}
\end{figure}

To assist in discerning the low-temperature behavior of $\Qi$ for on-defect devices, we reprint the on-defect data of Fig.~\ref{fig_Qi_vs_T} from the main text in Fig.~\ref{SI_fig_Qi_vs_T_ondefect}.

\subsection{Off-defect $\Qi(T)$ model}\label{SI_sect_off_defect_model}
Due to the placement of electrodes far from the mechanically active regions in off-defect designs, we do not expect the normal-metal, quasiparticle, or elastic loss channels of aluminum to affect the observed $\Qi(T)$. In Fig.~\ref{fig_Qi_vs_T} of the main text, the lack of any features in $\Qi(T)$ near the $T_c$ of aluminum, and the much higher $\Qi$ over a broad temperature range support this interpretation of off-defect devices. Instead, we model off-defect devices decaying via relaxation and resonant TLS loss channels. In the following discussion of TLS loss, we explicitly borrow the analysis and notation from Refs.~\cite{MacCabe2019,Behunin2016}, where more detailed derivations are provided.

For TLS with frequency $\omega_\text{TLS}$ coupled to a $d$-dimensional phonon bath in a bulk system, the average decay rate into the bath is given by~\cite{MacCabe2019}
\begin{align} \label{SI_eq_gamma_TLS}
    (\Gamma_{1,\text{TLS}})_{\text{ph},d} = \left( \frac{\bar{M}^2 \omega_{\text{TLS}}^d}{2\pi\hbar \bar{\rho}_\text{m} \bar{v}^{d+2}S_d}\right) \coth\left( \frac{\hbar \omega_{\text{TLS}}}{2k_B T} \right)\;,
\end{align}
where $\bar{M}$ is the average TLS transverse coupling potential to phonon bath modes, $\bar{\rho}_\text{m}$ is the bulk material mass density, $\bar{v}$ is the average acoustic velocity, and $S_d$ is the $(3-d)$ dimensional cross section. For a quasi-2D material, $S_d$ equals the film thickness, while for a quasi-1D material $S_d$ is the cross-sectional area of the thin beam. The conditions for reduced system dimensionality include~\cite{Behunin2016} (i) emitted phonons lower than the cutoff frequency of at least one system dimension, e.g. $\omega_\text{TLS}$ or thermal phonons $\omega_\text{th} \sim k_B T/\hbar$, or (ii) thermally active TLS are spaced further than at least one system dimension. 

Given the decay rate of TLS coupled to the phonon bath, it is then possible to calculate the resonator's relaxation damping $\gamma_{i,\text{rel}}$ from higher-order interactions with off-resonant TLS,~\cite{MacCabe2019}
\begin{align}
    \gamma_{i,\text{rel}} = \sum_{\text{TLS}} \left(\frac{2\bar{g}_l^2}{\omegap} \right) \left(\frac{\hbar\Gamma_{1,\text{TLS}}}{k_B T}\right) \text{sech}^2 \left( \frac{\hbar \omega_{\text{TLS}}}{2k_B T} \right)\;,
\end{align}
where $\bar{g}_l$ is the average longitudinal coupling rate between TLS and the resonator with frequency $\omegap$. This loss channel arises due to energy level modulation of the TLS by the resonator's oscillatory strain field, thereby displacing TLS from thermal equilibrium. Relaxation of TLS back to thermal equilibrium at rate $\Gamma_{1,\text{TLS}}$ effectively damps the resonator driving the TLS energy levels.

Substitution of Eq.~\ref{SI_eq_gamma_TLS} into the expression for $\gamma_{i,\text{rel}}$, and integration over a uniform TLS spectral density of states $n_0$ gives the quality factor $\Qirel$ due to relaxation TLS in a lower-dimensional bulk system,
\begin{multline} \label{SI_eq_Qirel_full}
    \Qirel^{-1}(T)  = n_0 \left( \frac{4\bar{g}_l^2}{\omegap^2} \right) \left( \frac{\bar{M}^2 }{2\pi\hbar\bar{\rho}_\text{m} \bar{v}^{d+2}S_d}\right) \left( \frac{k_B T}{\hbar}\right)^d \\
    \times\int_0^{\infty} x^d \text{csch} (x) dx\;,
\end{multline}
where we have converted the integral over all TLS frequencies to dimensionless units $x = \hbar \omega_{\text{TLS}}/k_B T$, and used the identity $\text{sech}^2(x/2)\coth(x/2) = 2\,\text{csch}(x)$. This reveals the $\Qirel^{-1}(T) \sim T^d$ temperature dependence on the effective phonon bath dimension, although $d$ is expected to change if conditions (i) and (ii) are satisfied for additional system dimensions as the temperature is further decreased. 

There are two important points on how Eq.~\ref{SI_eq_Qirel_full} applies to the off-defect device. First, for the relevant temperatures $T>1\,\text{K}$ where we observe relaxation TLS loss, we have $\omega_{\text{th}}/2\pi > 21\,\text{GHz}$, such that $\omega_{\text{th}} \gg \omegap$. At these frequencies, the defect site and mirror cell's density of states approximates the $d$-dimensional bulk, allowing us to neglect the discrete nature of the resonator's eigenmodes in our analysis. We also add that allowing for a weak energy dependence of the TLS density of states $n_0 \rightarrow n(\omega) \sim \omega^\mu$ would modify the observed temperature scaling $\Qirel^{-1}(T) \sim T^{d+\mu}$, with e.g. $\mu\simeq 0.3$ expected in glass~\cite{Behunin2016,Zolfagharkhani2005}. For our analysis, we simply treat $d$ as an effective bath dimension, neglecting the distinction between the effective geometrical dimension and the TLS density of states.

The phononic crystal resonators in this study are quasi-1D by design, however at elevated temperatures ($T>1\,\text{K}$) the system does not necessarily have a reduced phonon-bath dimensionality yet. For example, assuming a velocity $\bar{v} \simeq 6\times10^3\,\text{m/s}$ in bulk lithium niobate, we expect the wavelength of thermal phonons to exceed the thickness of the lithium niobate ($250\,\text{nm}$) near $T\simeq 1.1\,\text{K}$ and to exceed most feature sizes of the device ($150\,\text{nm} - 1\,\mu\text{m}$) by $T\simeq300\,\text{mK}$. However, below these temperatures resonant TLS already begin to dominate the resonator loss, preventing the observation of additional reduction in phonon-bath dimensionality. We also remark that the presence of the phononic bandgap will strongly affect the effective bath dimension seen by off-resonant TLS lying inside the fundamental and higher-order bandgaps. Observation of these effects would require further study.

To model the quality factor from relaxation TLS, we can lump all but the temperature dependence of Eq.~\ref{SI_eq_Qirel_full} into a term $\Qirel^{-1}(d)$ so that
\begin{align} \label{SI_eq_Qirel}
    \Qirel^{-1}(T) = \Qirel^{-1}(d) \times \left( T/T_0 \right)^d\;. 
\end{align}
Here, $T_0$ is a reference temperature for $\Qirel^{-1}(d)$, which can be arbitrarily chosen. We take $T_0 = 800\,\text{mK}$, approximately the temperature when resonant and relaxation loss are equal and $\Qi(T)$ peaks in Fig.~\ref{fig_Qi_vs_T}. Below this temperature, we presume that the system dimensionality should reduce further while approaching base temperature, in which case the simple power law $\Qirel^{-1}(T)\sim T^d$ breaks down. Therefore, with this reference temperature $\Qirel^{-1}(d)$ is the highest observed quality factor that can be attributed to relaxation TLS, representing a lower bound for the loss channel at $10\,\text{mK}$.

As in the on-defect case, we include the temperature dependence of resonant TLS via Eq.~\ref{SI_eq_Qires}, assuming a constant power-enhanced quality factor $\Qires(n)$. For the full fit to $\Qi(T)$ for the off-defect device in Fig.~\ref{fig_Qi_vs_T},
we have
\begin{align}
    \Qi^{-1}(T) = \Qires^{-1}(T) + \Qirel^{-1}(T)\;.
\end{align}
The best fit parameters and model constants are provided in Table~\ref{SI_table_Qi_T_params}, where the only fixed parameters are the measured resonant frequency $\omegap$, the average intra-cavity phonon number $\langle n \rangle$, and the (arbitrary) reference temperature $T_0$. The measurements of the off-defect device were performed at lower excitation levels, leading to less power-enhancement of $\Qires(n)$ compared to the on-defect device. However, both designs have approximately the same critical phonon number $n_c\simeq 2\times 10^2$ for saturation of resonant TLS.

\section{Mass sensitivity}\label{SI_sect_mass}

The nanomechanical resonators investigated in this study serve as extremely compact detectors of mass, with an approximately $0.25\,\mu\text{m}^3$ mechanically active region. During the cooldown process in our $4\,\text{K}$ tabletop cryostat, we observe that contaminants are adsorbed onto the surface of the mechanical resonators, leading to a frequency red-shift and degradation of quality factors $\Qi$. This mass flux conveniently allows us to explore the mass sensitivity of these devices both during the initial cooling and at base temperature.

\begin{figure}[t]
    \centering
    \includegraphics[width=85mm]{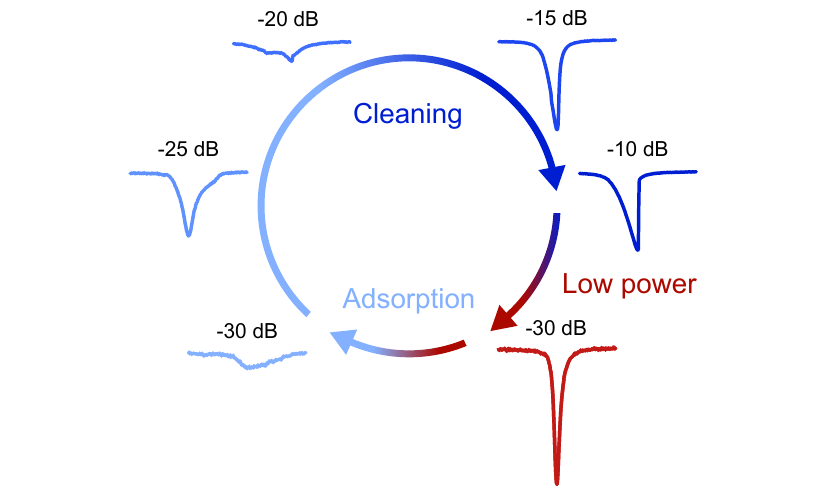}
    \caption{{\bf Cleaning cycle.} During the initial cryogenic cooling of the mechanical resonators in our $4\,\text{K}$ tabletop cryostat, the device response becomes non-lorentzian and quality factors $\Qi$ degrade due to adsorption of contaminants onto the resonator surface. The light blue trace ($-30\,\text{dB}$, bottom left) shows a typical reflection measurement after cooling to $4\,\text{K}$. In order to remove the adsorbate, we use a simple cleaning process in which the resonators are briefly measured using high microwave power. Increasing the measurement power (blue traces, going clockwise) to sufficiently high drive ($-10\,\text{dB}$) results in Duffing behavior and large mechanical displacement; an immediate return to low power (red, $-30\,\text{dB}$) yields significantly higher $\Qi$ than first measured and a return to the original resonance frequency, suggesting that the adsorbate is removed. Eventually the higher $\Qi$ will regress back to the initial low $\Qi$ state, although the timescale for this process is suspected to take days in our $4\,\text{K}$ tabletop cryostat.}
    \label{SI_fig_hyst}
\end{figure}

\begin{figure}[b]
    \centering
    \includegraphics[width=85mm]{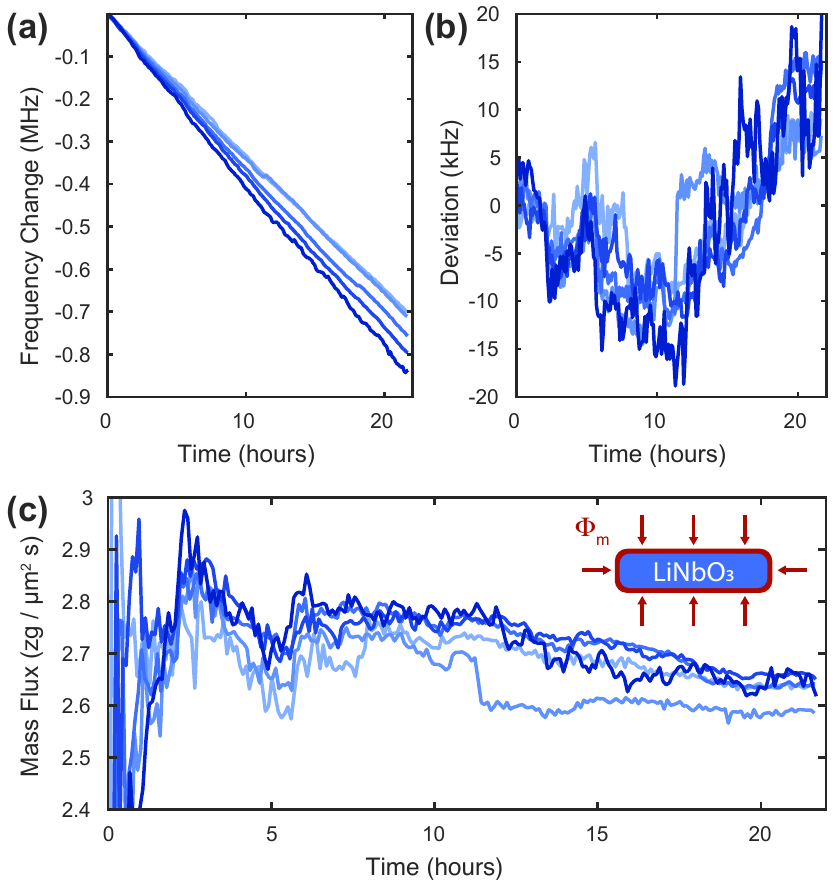}
    \caption{{\bf Mass sensitivity at $\mathbf{T =}$ 4\,K.} (a)~Change in five of the on-defect resonator's frequencies $\omegap$ over time due to adsorption of contaminants leaking into the $4\,\text{K}$ tabletop cryostat. Adsorption onto the mechanical defect sites change each resonator's effective mass, resulting in a frequency red-shift. (b)~Deviation in resonant frequency from a linear fit to the frequency red-shift time series. The high correlation coefficient $\sim0.8$ between simultaneously measured resonators suggests small variations in the adsorption rate on the timescale of hours. (c)~Estimate for the mass flux $\massflux$ over time. A simple adsorption model (inset) is used to model the accumulation of a thin layer which perturbs the effective mass of the resonator. The resonators converge to an estimate for $\massflux$ as the rate of frequency red-shift is more precisely known. In place of calibrating the detectors with a known mass flux, we use finite-element simulations to estimate the effective mass and displacement integrals for the simple adsorption model.}
    \label{SI_fig_mass_sensitivity}
\end{figure}

In order to initialize mass detection, we perform a simple cleaning process in which the resonators are briefly probed with high power immediately before the mass detection measurement (Fig.~\ref{SI_fig_hyst}). This has the effect of driving off adsorbates, resulting in an increase in $\Qi$ and resonant frequency to their maximum observed values. With this technique we are able to obtain consistent results for $\Qi$ both while cooling down and warming up the cryostat. After the resonators have been cleaned, contaminants slowly accumulate on the timescale of hours while operating at base temperature (Fig.~\ref{SI_fig_mass_sensitivity}). Presumably these contaminants are mostly nitrogen and oxygen from an atmospheric leak in our cryostat, as the mass flux persists over the course of weeks. In contrast to experiments in our $4\,\text{K}$ tabletop cryostat, we do not observe a detectable frequency red-shift of mechanical resonators in our dilution refrigerator (on the timescale of several weeks).

Here, we present a simple model from which we can estimate the mass flux incident on the mechanical resonators when cooled to $T = 4\,\text{K}$ in the presence of a small leak in our cryostat. For a localized mechanical mode with displacement $u(\mathbf{r})$ and density $\rho$, we can define the total kinetic energy as an integral over the mode's volume, $KE = \frac{1}{2}\int dV \dot{u}\rho \dot{u}$. In characterizing a mechanical resonator, it is useful to define the effective mass~\cite{Safavi2014book}
\begin{align}
    \meff \equiv \frac{\int dV \rho |u|^2}{\max |u|^2}\;,
\end{align}
which comes from expressing the mode's kinetic energy in the form of a harmonic oscillator, $KE = \frac{1}{2}\meff\omega^2 \max{|u|^2}$. In examining the fundamental shear modes of the mechanical resonators, finite-element simulations give estimated effective masses in the range of $\meff \simeq 490-750\,\text{fg}$ for the five on-defect resonators studied, and $\meff \simeq 440-670\,\text{fg}$ for the off-defect resonators (approximately 10\% lower). Simulations of this shear mode also show $\meff \simeq 0.42 \times m_{\text{tot}}$, where $m_{\text{tot}} = \int dV \rho$ is the total mass of the mechanically active region.

To incorporate adsorption into the model, we assume that the adsorbate uniformly covers the surface of the mechanical resonator (Fig.~\ref{SI_fig_mass_sensitivity}(c),~inset), thereby perturbing the effective mass of the mode. For an infinitesimal film of adsorbate with mass density $\lambda$, the effective mass becomes 
\begin{align}
    \meff &= \frac{\int dV \left(\rho + \lambda \delta(S)\right) |u|^2}{\max |u|^2} \\
    &= m_{\text{eff},0} + \lambda\, \frac{\int dS |u|^2}{\max |u|^2}\;,\label{SI_eq_meff}
\end{align}
where $\delta(S)$ is a delta function at the surface of the mechanical resonator, and $m_{\text{eff},0}$ is the unperturbed effective mass. The second term in Eq.~\ref{SI_eq_meff} represents a perturbation $\delta\meff$ of the effective mass, which can be detected by measuring changes $\delta\omegap$ in the original resonance frequency $\omega_{\text{p,0}}$,
\begin{align} \label{SI_eq_domega}
    \frac{\delta\omegap}{\omega_{\text{p,0}}} = -\frac{1}{2}\frac{\delta \meff}{m_{\text{eff},0}}\;.
\end{align}
Under the assumption that the adsorbate accumulates linearly in time, we can extract the mass flux $\massflux(t) = \lambda / t$ incident on the detector by combining Eqs.~\ref{SI_eq_meff}~and~\ref{SI_eq_domega}. For a frequency change $\delta\omegap$ measured over interval $\tau$, we obtain the mass flux incident on the device,
\begin{align}
    \massflux = -\frac{2\delta \omegap}{\omega_{\text{p,0}}\tau} \frac{\int dV \rho |u|^2}{\int dS |u|^2}\;.
\end{align}
Here, the surface and volume integrals can be calculated for the mechanical mode using numerical simulations, from which we estimate an approximate mass flux $\massflux \simeq 2.7\,\text{zg}/\mu\text{m}^2\cdot\text{s}$ in our $4\,\text{K}$ tabletop cryostat using the data in Fig.~\ref{SI_fig_mass_sensitivity}.

\end{document}